\documentclass[runningheads,a4paper]{llncs}

\newcommand{\ExtendedVersionOnly}[1]{#1}\newcommand{\ProceedingsVersionOnly}[1]{}

\usepackage[utf8]{inputenc}
\usepackage{amsmath}
\usepackage{amssymb}
\usepackage{esvect}
\usepackage{hyperref}
\usepackage{graphicx}
\usepackage{todonotes}
\usepackage{tabularx}
\usepackage{times}
\usepackage{listings}
\usepackage{xspace}
\usepackage[shell]{gnuplottex}
\usepackage{epstopdf}
\usepackage{xcolor,colortbl}
\usepackage{url} 

\hypersetup{bookmarksopen=true}

\newcommand{\ttlinline}[1]{\lstinline[language={ttl},basicstyle=\ttfamily\footnotesize,breaklines=false]|#1|}
\newcommand{\sparqlinline}[1]{\lstinline[language={sparql},basicstyle=\ttfamily\footnotesize,breaklines=false]|#1|}

\usepackage{color}
\definecolor{olivegreen}{rgb}{0.1,0.6,0.3}
\definecolor{grey}{rgb}{0.2,0.2,0.2}
\lstdefinelanguage{ttl}{
sensitive=true,
showstringspaces=false,
morecomment=[l][\color{grey}]{@},
morecomment=[l][\color{olivegreen}]{\#},
morestring=[b][\color{blue}]\",
keywordstyle=\color{cyan},
morekeywords={owl,rdf,rdfs,xml,xsd,dbo,dbp,dbprop,dbe,pat,guo,prov,void,changesets,dig,dcterms,item,update,snapshot}
}
\lstdefinelanguage{sparql}{
sensitive=true,
showstringspaces=false,
morecomment=[l][\color{grey}]{@},
morecomment=[l][\color{olivegreen}]{\#},
morestring=[b][\color{blue}]\",
keywordstyle=\color{cyan},
morekeywords={owl,rdf,rdfs,xml,xsd,dbo,dbp,dbprop,dbe,pat,guo,prov,void,changesets,dig,dcterms,item,update,snapshot}
}

\newcommand{\result}{\textsf{\footnotesize result}}
\newcommand{\execTime}{\textsf{\footnotesize execTime}}
\newcommand{\prevExecs}{\textsf{\footnotesize prevExecs}}
\newcommand{\prevExecsXXX}{\textsf{prevExecs}}  
\newcommand{\lastExec}{\textsf{\footnotesize lastExec}}
\newcommand{\lastExecXXX}{\textsf{lastExec}}  
\newcommand{\rank}{\textsf{\footnotesize rank}}
\newcommand{\jaccard}{\textsf{\footnotesize jaccard}}

\begin{document}

\mainmatter

\title{Scheduling Refresh Queries for Keeping Results from a SPARQL Endpoint Up-to-Date}
\titlerunning{Scheduling Refresh Queries for a SPARQL Endpoint}

\ProceedingsVersionOnly{%
\subtitle{(Short Paper)}
\nocite{ExtendedVersion}
}

\ExtendedVersionOnly{%
\subtitle{\vspace{-1mm}(Extended Version)%
\vspace{-6mm}\footnote{This document is an extended version of a paper published in ODBASE~2016~\cite{ProceedingsVersion}.}}
}

\author{
Magnus Knuth\inst{1}
\and Olaf Hartig\inst{2}
\and Harald Sack\inst{1}
}

\authorrunning{Magnus Knuth et~al.}

\institute{
Hasso Plattner Institute, University of Potsdam, Germany\\
\email{\{magnus.knuth|harald.sack\}@hpi.de}
\\[2mm] 
\and
Dept.\ of Computer and Information Science (IDA), Link\"oping University, Sweden\\
\email{olaf.hartig@liu.se}
}

\maketitle

\begin{abstract}
Many datasets change over time. As a consequence, long-running applications that cache and repeatedly use query results obtained from a SPARQL endpoint may resubmit the queries regularly to ensure up-to-dateness of the results. While this approach may be feasible if the number of such regular refresh queries is manageable, with an increasing number of applications adopting this approach, the SPARQL endpoint may become overloaded with such refresh queries.
	A
more scalable approach would be to use a middle-ware component at which the applications register their queries and get notified with updated query results once the results have changed. Then, this middle-ware can schedule the repeated execution of the refresh queries without overloading the endpoint.
In this paper, we study the problem of scheduling refresh queries for a large number of registered queries by assuming an overload-avoiding upper bound on the
	length of a regular time slot available for testing refresh queries.
We investigate a variety of scheduling strategies and compare them experimentally in terms of time slots needed before they recognize changes and number of changes that they miss.
\end{abstract}

\section{Introduction}
\label{sec:introduction}


	Many datasets on the Web of Data reflect data related to current events or ongoing activities. Thus, such datasets
are dynamic and evolve over time \cite{kafer-t-2013-dyldo}.
As a consequence, query results
	that have been
obtained from a SPARQL endpoint may become outdated.
Therefore, long-running applications that cache and repeatedly use query results have to resubmit the queries regularly to ensure up-to-dateness of the~results.

	There would be no need for such regular tests
if SPARQL endpoints would provide information about dataset modifications. 
	There exist manifold approaches
for providing such information. Examples are cache validators for SPARQL requests~(using HTTP header fields such as \texttt{\footnotesize Last-Modified} or \texttt{\footnotesize ETag})~\cite{williams-g-2011-enabling} 
and publicly available dataset update logs~(as provided by
	DBpedia Live at {\footnotesize \url{http://live.dbpedia.org/changesets/}}).
Unfortunately, existing SPARQL endpoints rarely support such approaches~\cite{kjernsmo-k-2015-survey}, nor is
	update information
provided in any other form by the dataset providers.
The information needed has to be generated by the datastore underlying the SPARQL endpoint or by dataset wrappers that exclusively control all the updates applied to the dataset, which is often not possible, e.g.\ in the case of popular RDB2RDF servers, as they typically work as one-way RDF exporters.
Without information about dataset modifications and changes from dataset side, the only viable alternative is to re-execute the respective SPARQL queries and check whether the obtained results have changed. 
This approach is feasible only if the number of such regular refresh queries is manageable. With an increasing number of applications adopting this approach, the SPARQL endpoint might become overloaded with the refresh queries. 
A~more scalable approach would be to use a middle-ware component at which the applications register their queries and get notified updates once the query results have changed.
Then, this middle-ware is able to schedule the repeated execution of the refresh queries without risking to overload~the~endpoint.

A main use case of such a middle-ware is the sparqlPuSH approach to provide a notification service for data updates in RDF stores \cite{passant-a-2010-sparqlpush}. 
	sparqlPuSH
relies on SPARQL queries and tracks changes of the result sets that then are published as an RSS feed and broadcasted via the PubSubHubbub protocol \cite{fitzpatrick-b-2010-pubsubhubbub}. 
However,
the existing implementation of sparqlPuSH is limited to the
	particular use case of micro-posts and circumvents the problem of detecting changes by expecting dataset updates to be performed via the sparqlPuSH interface%
		~\cite{knuth-m-2015-towards}%
	.
To generalize the idea of sparqlPuSH scheduling the re-eval\-u\-a\-tion of SPARQL queries has been identified as an unsolved research~problem~\cite{knuth-m-2015-towards}.


In this paper, we study this problem of scheduling refresh queries for a large number of registered SPARQL queries; as an overload-avoiding constraint we assume an upper bound on
	the length of time slots during which sequences of refresh queries can be run.
We investigate various scheduling strategies and compare them experimentally%
. 
For our experiments, we use a highly dynamic real-world
	dataset over a period of three months, in combination with a dedicated set of queries. The dataset~(DBpedia~Live) comprises all real-time changes in the Wikipedia that are relevant for DBpedia.

The main contributions of the paper are an empirical evaluation of a corpus of real-world SPARQL queries regarding result set changes on a dynamic dataset and an experimental evaluation of different query re-evaluation strategies.
Our experiments show that the change history of query results is the main influential factor, and scheduling strategies based on 
	the extent of previously recognized changes (dynamics)
and 
	an adaptively allocated maximum lifetime for individual query results 
	provide the best performances.

The remainder of the paper is structured as follows: 
Sec.~\ref{sec:related} discusses related work.
Sec.~\ref{sec:definitions} provides definitions and prerequisites.
These are needed for
Sec.~\ref{sec:strategies} which introduces the scheduling strategies used for the experiments.
Sec.~\ref{sec:setup} describes the experimental setup, including the dataset and queryset that we used and the applied evaluation metrics.
Sec.~\ref{sec:experiments} and Sec.~\ref{sec:conclusions} present the experimental results and discuss them, respectively. 
Sec.~\ref{sec:outlook} concludes the paper with an outlook on ongoing and future~work.

\section{Related Work}
\label{sec:related}


A variety of existing applications is related to change detection of query results on dynamic RDF datasets, such as (external) query caching~\cite{martin-m-2010-improving}, partial dataset update~\cite{endris-k-2015-interest}, as well as notification services~\cite{passant-a-2010-sparqlpush}.
However, even though Williams and Weaver show how the \texttt{\footnotesize Last-Modified} date can be computed with reasonable modifications to a state-of-the-art SPARQL processor~\cite{williams-g-2011-enabling}, working implementations are rare.
In fact, Kjernsmo has shown in an empirical survey that only a miniscule fraction of public SPARQL endpoints actually support caching mechanisms on a per-query basis~\cite{kjernsmo-k-2015-survey}. 

To overcome this lack of direct cache indicators, alternative approaches are required to recognize dataset updates.
The most common approach is to redirect updates through a wrapper that records all changes~\cite{martin-m-2010-improving,passant-a-2010-sparqlpush}. However, this approach is not applicable for datasets published by someone else.
If data publishers provide information on dataset updates, this information can be analyzed. For instance, Endris et al.\ introduce an approach to monitor the changesets of DBpedia~Live for relevant updates~\cite{endris-k-2015-interest}~(such a changeset is a log of removed and inserted triples).
Tools for dataset update notification, such as \emph{DSNotify}~\cite{popitsch-n-2011-dsnotify} and \emph{Semantic Pingback}~\cite{tramp-s-2010-weaving}, are available but extremely rarely deployed.
Further hints for possible changes may be obtained from metadata about datasets%
	; for instance, the
DCAT recommendation suggests to~use \ttlinline{dcterms:modified} or \ttlinline{dcterms:accrualPeriodicity} to
	describe update frequencies of a dataset.
\footnote{\url{http://www.w3.org/TR/vocab-dcat/}}

	Since the aforementioned cache indicators and hints for change detection are missing almost entirely in practice,
we rely on re-execution of queries. Apparently, such an approach causes overhead in terms of additional network traffic and server load. In order to reduce this overhead we investigate effective scheduling strategies in this paper.
A similar investigation in the context of updates of Linked Data has been presented by Dividino~et~al.~\cite{dividino-r-2015-strategies}. The authors show that change-aware strategies
	are suitable
to keep local \emph{data caches} up-to-date.
	We
also evaluate a strategy adopted from
	Dividino~et~al.'s
\emph{dynamicity} measure. We observe that, in our context, this strategy performs well for highly dynamic queries, but it is prone to starvation for less dynamic~queries.

	Query result caches are also used for database systems
		where the main use case is to enhance the scalability of backend databases
	for dynamic da\-ta\-base-driven websites.
	The most prominent system is \emph{Memcached}\footnote{\url{http://www.memcached.org/}} which supports the definition of an expiration time for individual cache entries, as well as local cache invalidation, e.\,g. when a client itself performs an update. Consequently, updates from other sources cannot be invalidated.
	More sophisticated systems, such as the proxy-based query result cache \emph{Ferdinand} \cite{garrod-c-2008-queryresult}, use update notifications to invalidate local caches. To determine the queries that are affected by an update it is necessary to solve the query-update dependence problem~\cite{levy-a-1993-queries}. This process demands access to the dataset updates, which, as said, are not available in the general case for externally published~Linked~Datasets.

\section{Preliminaries}
\label{sec:definitions}

In this paper we consider a dynamic dataset, denoted by $\mathcal{D}$, that gets updated continuously or in regular time intervals. We assume a sequence $\vv{\mathcal{T}} = (t_1, t_2, \dots, t_n)$ of consecutive points in time
at which the dataset constitutes differing revisions%
.
Additionally, we consider a finite set~$Q$ of SPARQL queries.
Then, for every time point $t_i$ in $\vv{\mathcal{T}}$ and for every query~$q \in Q$, we write $\result(q,i)$ to denote the query result that one would obtain when executing $q$ over $\mathcal{D}$ at $t_i$.
Furthermore, let $C_i \subseteq Q$ be the subset of the queries whose result at $t_i$ differs from the result at the previous time point $t_{i-1}$, i.e.,
$$
C_i = \big\lbrace q \in Q \mid \result(q,i) \neq \result(q,i-1) \big\rbrace .
$$
The overall aim is to identify a greatest possible subset of $C_i$ at each time point~$t_i$. A~trivial solution to achieve this goal would be to execute all queries from $Q$ at all time points. While this exhaustive approach may be possible for a small set of queries, we assume that the size of $Q$ is large enough for the exhaustive approach to seriously stress, or even overload, the query processing service. Therefore, we consider an additional politeness constraint that any possible approach has to satisfy. For the sake of simplicity, in this paper we use as such a constraint an upper bound on the size of the time slots within which approaches are allowed to execute a selected sequence of queries for the different time points. Hereafter, let $K_\mathsf{maxExecTime}$ be this upper bound, and, for any possible approach, let $E_i \subseteq Q$ be the~(refresh) queries that the approach executes in the time slot for time point~$t_i$. Hence, if we let $\execTime(q,i)$ denote the time for executing $q$ over the snapshot of $\mathcal{D}$ at~$t_i$, then for all past time points we have 
$$K_\mathsf{maxExecTime} \geq \sum_{q \in E_i} \execTime(q,i) .
$$

To select a sequence of queries to be executed within the time slot for a next time point, the approaches may use any kind of information obtained by the query executions performed during previous time slots for earlier time points.
  For instance, to select the sequence of queries for a time point $t_i$, an approach may use any query result $\result(q,j)$ with $j < i$ and $q \in E_j$, but it cannot use any $\result(q'\!,j')$ with $q' \notin E_{j'}$ or with $j' \geq i$.

As a last preliminary, in the definition of some of the approaches that we are going to introduce in the next section we write $\prevExecs(q,i)$ to denote the set of all time points for which the corresponding approach executed query $q \in Q$ before arriving at time point $t_i$;
	i.e.\
$
\prevExecs(q,i) = \lbrace j < i \mid q \in E_j \rbrace.
$
In addition, we write $\lastExec(q,i)$ to denote the most recent of these time points, i.e.\
$
\lastExec(q,i) = \max\bigl( \prevExecs(q,i) \bigr) .
$




\section{Scheduling Strategies}
\label{sec:strategies}

	This section presents
the scheduling strategies implemented for our evaluation. We begin
by introducing features that
	may affect
the behavior of such strategies.

Typically, dataset providers do not offer any mechanism to inform clients about data updates, neither whether the data has changed nor to what extent. Therefore, we focus on scheduling strategies that are dataset agnostic, i.\,e. strategies that do not assume information about what has changed since the last query execution.
Hence, all features that such a strategy can exploit to schedule queries for the next refresh time slot originate from
(a)~the queries themselves,
(b)~an initial execution of each query, and
(c)~the ever growing history of successful executions of the queries during previous time~slots.

Given these constraints, we have implemented different scheduling policies using the following features:
\begin{itemize}
  \item \emph{Age} describes the actual time passed since the last query execution%
	.
  
  \smallskip
  
  \item \emph{Estimated execution time} is computed from the median query execution time over the last query executions and corresponds to the politeness constraint $K_\mathsf{maxExecTime}$.
  
  \smallskip
  
  \item \emph{Change Rate} indicates ``how often'' a query result has changed. It is derived from the recognition of result changes within the last query executions.
  
  \smallskip
  
  \item \emph{Change Dynamics} indicates ``to what extent'' a query result has changed. It is an aggregation of result changes over the last query executions%
	~\cite{dividino-r-2014-changes}. We compute this
		metric
	by using the \emph{Jaccard distance} between known subsequent results.
\end{itemize}

We have implemented seven scheduling policies known from the literature%
	. We classify them into
two groups: \emph{non-selective} and \emph{selective} policies.
	By using a \emph{non-selective} scheduling policy, potentially all registered
queries are evaluated according to
	a
ranking order until the execution time limit~($K_\mathsf{maxExecTime}$) has been reached. For every time point $t_i$ in $\vv{\mathcal{T}}$\!, a new ranking for all queries is determined.
The queries are ranked in ascending order using a ranking function $\rank(q,i)$. In a tie situation,
	the decision is made based on the age of the query, and finally the query id.

\begin{description}
  \item [Round-Robin (RR)] treats all queries equal disregarding their change behavior and execution times. It
  executes the queries for which the least current result
  is available.
  \begin{equation}
  \rank_{RR}(q,i) = \frac{1}{i- \lastExecXXX(q,i)}
  \end{equation}
  \item [Shortest-Job-First (SJF)] prefers queries with a short estimated runtime~(to execute as many queries per time slot as possible). The runtime is estimated using the median value of runtimes from previous executions. Additionally, the exponential decay function $e^{-\lambda(i- \lastExecXXX(q,i))}$ is used as an aging factor to prevent starvation.
  \begin{equation}
  \rank_{SJF}(q,i) = e^{-\lambda (i- \lastExecXXX(q,i))} \mathrm{median}_{j \in \prevExecsXXX(q,i)}\bigl( \execTime(q,j) \bigr)
  \end{equation}
  \item [Longest-Job-First (LJF)]
  	uses the same runtime estimation and aging as SJF but prefers long estimated runtimes, assuming such queries are more likely to produce a result.
  \begin{equation}
  \rank_{LJF}(q,i) = \frac{ e^{-\lambda (i- \lastExecXXX(q,i))} }{ \mathrm{median}_{j \in \prevExecsXXX(q,i)}\bigl( \execTime(q,j) \bigr) }
  \end{equation}
  \item [Change-Rate (CR)] prioritizes queries that have changed most frequently in the past%
  . A decay function $e^{-\lambda t}$ is used to weight the change added by its respective age.
  \begin{align}
  \rank_{CR}(q,i) &= \sum_{j \in \prevExecsXXX(q,i)} \!\left( e^{-\lambda (i-j)} * \mathsf{change}(q,i)\right),
  \\
  \text{where:} \quad \mathsf{change}(q,i) &= \begin{cases} 1 & \text{if } \result(q,j) \neq \result(q,\lastExec(q,j) ), \\ -1 & \text{else}. \end{cases}
  \end{align}
  \item [Dynamics-Jaccard (DJ)] has been proposed as a best-effort scheduling policy for data\-set updates~\cite{dividino-r-2015-strategies}. Here, for \texttt{\footnotesize DESCRIBE} and \texttt{\footnotesize CONSTRUCT} queries we compute the \emph{Jaccard distance} on RDF triples, and on the query solutions for \texttt{\footnotesize SELECT} queries. For \texttt{\footnotesize ASK} queries, the distance is either $0$ or $1$.
  \begin{align}
  \rank_{DJ}(q,i) &= \sum_{j \in \prevExecsXXX(q,i)} \!\left(e^{-(i-j)} * \jaccard(q,j) \right)
  \\
  \text{where:} \quad
  \jaccard(q,j) &= 1 - \frac{\bigl| \result(q,j) \cap \result(q,\lastExec(q,j)) \bigr|}{\bigl| \result(q,j) \cup \result(q,\lastExec(q,j)) \bigr|}
  \end{align}
\end{description}

Instead of ranking all queries, the \emph{selective} scheduling policies select a (potentially
	ranked)
subset of queries for evaluation at a given point in time $t_i$.
Queries
	from this subset that do not get evaluated due to
the execution time limit~($K_\mathsf{maxExecTime}$) are privileged in the next time slot~$t_{i+1}$.

\begin{description}
  \item [Clairvoyant (CV)]
  	is assumed to have full knowledge of all query results at every point in time and, thus, is able to determine the optimal schedule.
  \item [Time-To-Live (TTL)] determines specific time points when a query should be executed. To this end, each query is associated with a value indicating a time interval after which the query needs to be re-evaluated. After an evaluation, if the query result has changed, this time-to-live value is divided in half or, alternatively, reset to the initial value of $1$; if the result did not change, the value is doubled up to a fixed maximum value~($max$). We investigate different values as maximum time-to-live.
\end{description}

\section{Experimental Setup}
\label{sec:setup}


We evaluated the performances of the scheduling strategies experimentally.
In this section, we explain the test setup. The setup consists of a highly dynamic dataset and a corresponding set of SPARQL queries. The individual characteristics of the dataset and the query set are analyzed in detail, before we focus on the evaluation metrics.

\subsection{Dataset}
\label{sec:dataset}

For our experiments we use the
\emph{DBpedia~Live} dataset \cite{hellmann-s-2009-dbpedialive} because it provides continuous fine-grained changesets, which are necessary to reproduce a sufficient number of dataset revisions.
Moreover, while \emph{DBpedia~Live} and \emph{DBpedia} share the same structural backbone -- both make use of the same vocabularies and are extracted from English Wikipedia articles -- the main difference is that the real-time extraction of \emph{DBpedia~Live} makes use of different article revisions. Therefore, queries for \emph{DBpedia} can be expected to work alike for \emph{DBpedia~Live}, as we show in Sec.~\ref{sec:queries}.

We selected
	the three-months period August--October 2015
for replaying the changesets, starting from a dump of June
	2015~({\footnotesize \url{http://live.dbpedia.org/dumps/dbpedia_2015_06_02.nt.gz}})
applied with subsequent updates for June and July 2015.
After each fully~replayed hour, we collect
dataset statistics and execute the full query set. All statistics and results are
	recorded in a
database for the actual evaluation of the scheduling~strategies.

\begin{figure}[htp]
    \centering
    \input{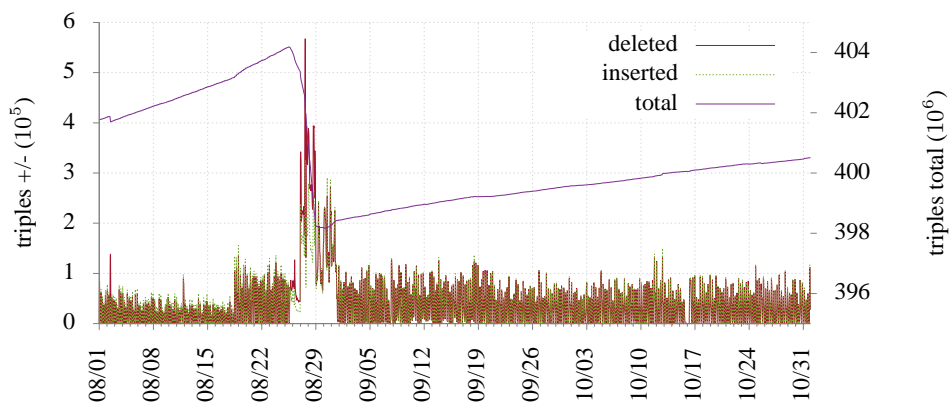}
    \vspace{-4em}
    \caption{Revision statistics}
    \label{fig:revstats}
\end{figure}

As shown in Fig.~\ref{fig:revstats}, the dataset contains between 398M and 404M triples.
The dataset changes are not homogeneous: starting from 08/18 we observe an increased number of triple updates, and from 08/27 to 08/31
	there have been exceptionally many insertions and even more deletions%
%
	~(the reason for this pattern could not be revealed from the changesets).
In total we have 2,208 hourly updates for
	our three-months period
(92 days * 24 hours), and there are 437 revisions (hours) without any
changes.

\subsection{Queries}
\label{sec:queries}

To perform SPARQL query executions on a dynamic dataset it is essential to use queries that match the dataset. 
We use a set of real-world queries from the \emph{Linked SPARQL Queries dataset} (LSQ) \cite{saleem-m-2015-lsq} which contains 782,364 queries for DBpedia. Though the queries originate from the year 2010 (DBpedia 3.5.1), they still match the current dataset structure.
We randomly selected 10,000
queries from
	LSQ
after filtering out those having a runtime of more than 10 minutes or producing parse or runtime errors.
The query set contains 11~\texttt{\footnotesize DESCRIBE}, 93~\texttt{\footnotesize CONSTRUCT}, 438~\texttt{\footnotesize ASK}, and 9458~\texttt{\footnotesize SELECT} queries%
	, and is available at {\footnotesize \url{https://semanticmultimedia.github.io/RefreshQueries/data/queries.txt}}.
DBpedia~Live changes gradually, but obviously the structural backbone of DBpedia remains. As a result, 4,423 out of our 10,000 queries deliver a non-empty query result on the first examined revision (4,440 over all examined revisions).

\begin{figure}[htp]
    \centering
    \input{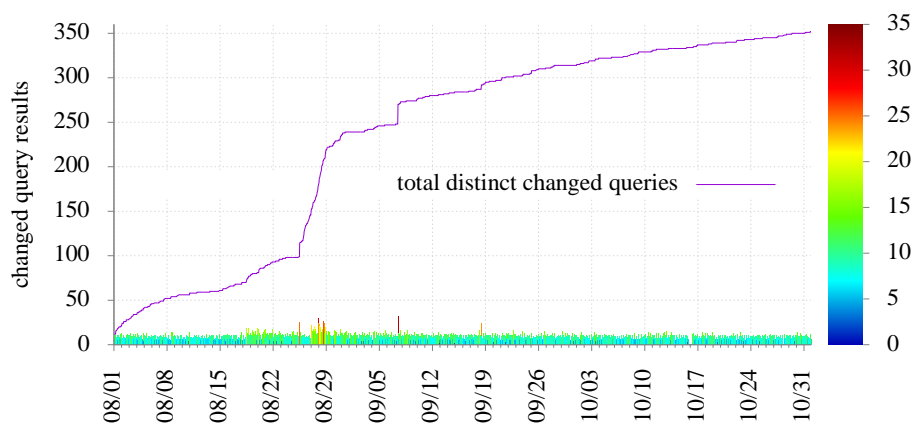}
    \vspace{-4em}
    \caption{Queries with update per revision (bars) and distinctly aggregated (line)}
    \label{fig:rschange}
\end{figure}

We consider a result as \emph{changed}, if it is not isomorphic to the result returned for this query in the previous evaluation. For queries having the \texttt{\footnotesize ORDER BY} feature we also check for an equal bindings sequence. If \texttt{\footnotesize ORDER BY} is not used in the query, the binding order is ignored as SPARQL result sets are then expected in no specific order~\cite{harris-s-2013-sparql}.

Concerning the \emph{result changes} (cf.\ Fig.~\ref{fig:rschange}) we observe that only a small fraction~of the queries is affected by the dataset updates~(up to 32~queries per revision, 352~queries within all revisions). Furthermore, by the continuously increasing number of total distinct queries with changed result, we observe that query results may also change after being constant for a long time.
Periods with higher data update frequencies~(e.g., from~08/27 to~08/31) can be identified also as periods with more query result~changes.

\begin{figure}[htb]
    \centering
    \input{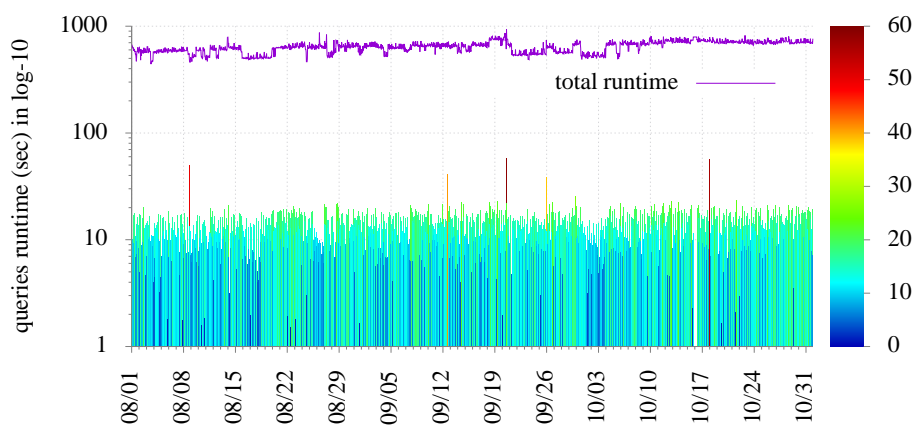}
    \vspace{-4em}
    \caption{Runtime of queries total vs. with update}
    \label{fig:rschangeruntime}
\end{figure}

As illustrated in Fig.~\ref{fig:rschangeruntime}, the overall runtime of all queries per revision ranges from 440 to 870 seconds, whereas the runtime for affected queries ranges 
up to 50.1 seconds~(consuming at maximum 8.9\,\% of the total runtime).

\begin{figure}[htb]
    \centering
    \input{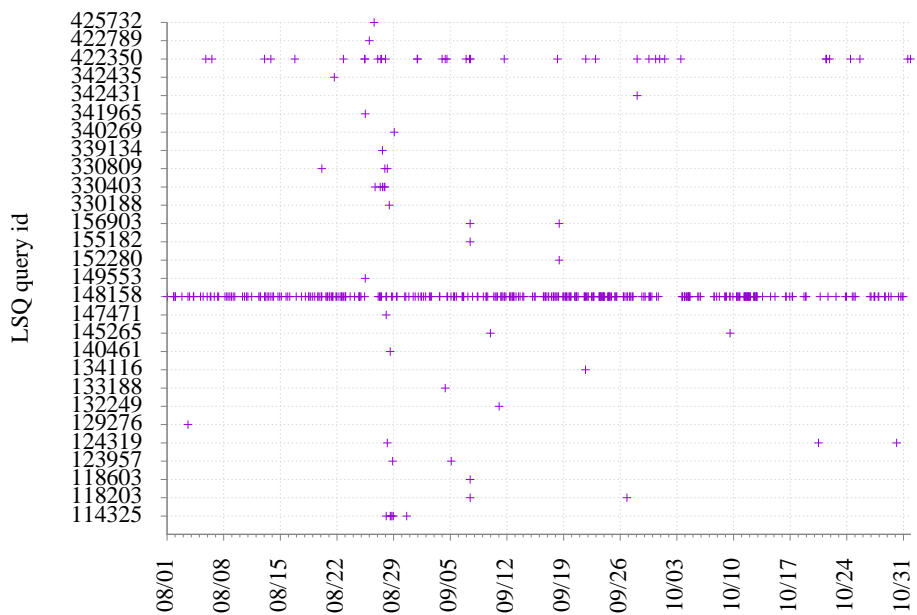}
    \vspace{-4em}
    \caption{Result changes per query (examples)}
    \label{fig:rschangedist}
\end{figure}

Fig.~\ref{fig:rschangedist} shows the individual time points the query result changes for a subset of the analyzed query set\footnote{Details on the individual queries can be retrieved from the LSQ dataset, accessible at \url{http://lsq.aksw.org/page/res/DBpedia-q<QUERY_ID>}.}.
The majority (191) of the 352 queries affected by the dataset updates change exactly once, 38 queries change twice. The result of the query\footnote{Shortened, find the original query at \url{http://lsq.aksw.org/page/res/DBpedia-q312238}.}\\
\sparqlinline{SELECT ?res ?v WHERE \{ ?res dbo:abstract ?v \} ORDER BY ?res ?v}\\
changes most often with 1,765 times.
We can recognize that the query results change in very irregular intervals with a high variation between the individual queries. The average interval between subsequent changes is 27.6 hours (standard deviation 145.6 hours) for the 352 queries which are affected by dataset updates.

The dataset replay and the query executions have been performed on a 48-core Intel(R)~Xeon(R)~CPU~E5-2695~v2~@2.40GHz 
using the AKSW Jena SPARQL API
\footnote{\url{https://github.com/AKSW/jena-sparql-api}} and an OpenLink Virtuoso Server~07.10 
with 32GB reserved~RAM.

\subsection{Publication of Experimental Data}
\label{sec:datapub}

We provide the data gathered from the experiments in form of a MySQL database dump
and an RDF dump
with the query executions as planned by the evaluated strategies\footnote{Both datasets are available at \url{https://semanticmultimedia.github.io/RefreshQueries/}}.

The database dump includes the plain results of all query executions, while the RDF dataset refers to their SHA256 hash values. The RDF dataset applies the LSQ vocabulary\footnote{\url{https://github.com/AKSW/LSQ/blob/gh-pages/LSQ_Vocab.rdf}}. We extended the vocabulary to describe relevant metadata such as the delay and the missed updates of individual query executions.

\subsection{Evaluation metrics}
\label{sec:metrics}

An ideal scheduling strategy should satisfy a number of requirements:
\begin{itemize}
\item \emph{Effectiveness}: It should only evaluate queries that have changed, which reduces unnecessary load to the SPARQL endpoint.

\smallskip

\item \emph{Efficiency}: It should evaluate queries that have changed as soon as possible, which reduces the out-of-date time and helps to not miss result changes.

\smallskip

\item \emph{Avoid starvation}: Results of queries that are susceptible to change~(i.e., there is no reason to believe the query will always produce the same result) may change at any point in time even if the results have been constant so far. It should be ensured that such queries are executed at some point.
\end{itemize}

To compare the query execution strategies we simulate their query selection with different configurations over all 2,208 dataset revisions ($t_1, \dots, t_{2208}$).
The initial query results $\lbrace \forall q \in Q: \result(q,0) \rbrace$ for $t_0 < 08/01$ are available to every scheduling strategy right from the start. We compute the following key metrics:

\begin{description}
  \item [Total query executions] number of query executions performed. 

\smallskip

  \item [Irrelevant executions] query executions without recognizing a change, equals to the total number of executions minus the relevant ones. Irrelevant executions create unnecessary load to the endpoint and reduce the \emph{effectiveness}.

\smallskip

  \item [Relevant executions] query executions where a change could be detected compared to the last execution, i.\,e. there was at least one result change since the execution; if there was more than a single change, these updates are counted as missed.

\smallskip

  \item [Effectivity] the ratio of relevant query executions to total executions.

\smallskip

  \item [Absolute delay] time between the optimal and actual re-execution $(q,i)$, summed
  over all queries, which
		allows to measure
  the overall \emph{efficiency} of the scheduling strategy.

\smallskip

  \item [Maximum delay] the longest delay for an individual query execution determines the maximum out-of-date time to be expected from the scheduling strategy for an individual query result. Overly long out-of-date times indicate a \emph{starvation} problem.

\smallskip

  \item [Absolute miss] number of changes that
  	are
  recognized, summed
  over all queries.

\smallskip

  \item [Maximum miss] the maximum number of missed result updates across all queries.
\end{description}

\newcommand{\clG}{\cellcolor{green!25}}  
\newcommand{\clg}{\cellcolor{green!10}}  
\newcommand{\clR}{\cellcolor{red!25}}  
\newcommand{\clr}{\cellcolor{red!10}}  

\section{Experimental Results}
\label{sec:experiments}

	We have conducted the experiment for three different values of
$K_\mathsf{maxExecTime}$: 10~sec, 50~sec, and 1,000~sec.
This variation of the upper bound execution time allows us to pretend different workloads: As we assume a fixed one-hour interval stepping with 10,000 queries, the workload can be scaled in terms of the number of queries and the time interval, respectively.
In the following we present the results for each configuration. The metrics as introduced in Sec.~\ref{sec:metrics} are listed in tabular form. The two best and worst achieved results per metric are highlighted in shades of green and red, respectively.

Table~\ref{tab:1000sec} shows the results for
	$K_\mathsf{maxExecTime}=$
1,000~sec, which, for our query set, is equivalent to unlimited runtime; that is, all queries could be executed for every revision.

Consequently, the theoretically optimal CV policy has no misses and delay, and executes only relevant queries.
In contrast, as the non-selective scheduling policies
	(RR/ SJF/LJF/CR/DJ)
execute all queries and therefore detect all relevant changes, they execute a massive amount of irrelevant queries as overhead, resulting in a low effectivity.

	The selective TTL policy reduces the number of query executions
effectively, and
	more updates are detected by resetting a query's time-to-live when a change has been detected.
The best performing configuration tested (TTL$_{{max}=32,reset}$) detects 81\,\% of all changes (12,311 of 15,256) while performing only 3.4\,\% of the query executions compared to the non-selective policies~(738,566 vs.\ 22,064,744). And still, TTL$_{{max}=256}$ detects 75\,\% (11,459) with 0.75\,\% query executions (154,185).
The reduced query overhead comes at the expense of more delay and in particular higher maximum delay times.

\begin{table}[ht]
\centering
\small
\caption{Config $K_\mathsf{maxExecTime}=1000sec$}
\label{tab:1000sec}
\begin{tabular}{l||r|r|r|r||r|r||r|r}
                       & \begin{tabular}{@{}c@{}} total qe \end{tabular}
                       & \begin{tabular}{@{}c@{}} irrelevant \end{tabular}
                       & \begin{tabular}{@{}c@{}} relevant \end{tabular}
                       & \begin{tabular}{@{}c@{}} eff. (\%) \end{tabular}
                       & \begin{tabular}{@{}c@{}} abs\\ delay \end{tabular}
                       & \begin{tabular}{@{}c@{}} max\\ delay \end{tabular}
                       & \begin{tabular}{@{}c@{}} abs\\ miss \end{tabular} 
                       & \begin{tabular}{@{}c@{}} max\\ miss \end{tabular} \\ \hline
CV                     &     15,256 &               0 &     15,256 &      100 &           0 &         0 &          0 &        0 \\ \hline
RR/SJF/LJF/CR/DJ       & 22,080,000 &\clR  22,064,744 &\clG 15,256 &\clR  .07 &\clG       0 &\clG     0 &\clG      0 &\clG    0 \\
TTL$_{{max}=32}$       &    744,565 &         732,685 &     11,880 &\clr 1.60 &      26,866 &\clg    31 &      3,376 &\clr   19 \\
TTL$_{{max}=32,reset}$ &    750,877 &\clr     738,566 &\clg 12,311 &     1.64 &\clg  23,492 &\clg    31 &\clg  2,945 &\clr   19 \\
TTL$_{{max}=64}$       &    405,175 &         393,507 &     11,668 &     2.88 &      40,747 &        63 &      3,588 &\clr   19 \\
TTL$_{{max}=128}$      &    245,246 &\clg     233,683 &\clr 11,563 &     4.71 &\clr  61,639 &\clr   127 &\clr  3,693 &\clr   19 \\
TTL$_{{max}=128,reset}$&    252,714 &         240,550 &     12,164 &\clg 4.81 &      53,655 &\clr   127 &      3,092 &\clr   19 \\
TTL$_{{max}=256}$      &    165,644 &\clG     154,185 &\clR 11,459 &\clG 6.92 &\clR  86,202 &\clR   255 &\clR  3,797 &\clr   19 
\end{tabular}
\end{table}

Table \ref{tab:50sec} shows the evaluation results for a runtime limitation of 50~seconds, which corresponds roughly to the maximum runtime needed for executing all relevant queries of the query set (cf.\ Sec.~\ref{sec:queries}).
The CV policy has no miss, but it cannot execute all queries on time; instead, it delays three relevant executions for one revision each.

As expected, SJF has most and LJF has least query executions given an execution time limitation, because short respectively long running queries are preferred. As the decay factor $\lambda$ is increased, in both cases the number of executed queries tends towards RR. Nevertheless, none of both strategies outperforms RR regarding relevant query executions, delay, or number of misses.
The change rate based policies (CR) demonstrate that the result history is a good indicator and a significant number of changes was detected: 92.9\,\% for CR$_{\lambda=0.0}$ and 66.7\,\% for CR$_{\lambda=0.5}$.
The dynamicity-based policy (DJ) detects by far the most result updates (99.7\,\%) and produces the least delay; the effectiveness is above CR.
The TTL configurations show comparable results to the 1000~seconds runtime limitation, i.\,e. the number of total query executions, detected changes, and the delay remain relatively stable with the 50~seconds limit. Again, we see most result updates are detected by the TTL$_{{max}=32,reset}$ configuration.

\begin{table}[ht]
\centering
\small
\caption{Config $K_\mathsf{maxExecTime}=50sec$}
\label{tab:50sec}
\begin{tabular}{l||r|r|r|r||r|r||r|r}
                       & \begin{tabular}{@{}c@{}} total qe \end{tabular}
                       & \begin{tabular}{@{}c@{}} irrelevant \end{tabular}
                       & \begin{tabular}{@{}c@{}} relevant \end{tabular}
                       & \begin{tabular}{@{}c@{}} eff. (\%) \end{tabular}
                       & \begin{tabular}{@{}c@{}} abs\\ delay \end{tabular}
                       & \begin{tabular}{@{}c@{}} max\\ delay \end{tabular}
                       & \begin{tabular}{@{}c@{}} abs\\ miss \end{tabular} 
                       & \begin{tabular}{@{}c@{}} max\\ miss \end{tabular} \\ \hline
CV                     &     15,256 &             0 &     15,256 &      100 &           3 &         1 &          0 &         0 \\ \hline
LJF$_{\lambda=0.5}$    &    977,922 &       974,512 &\clR  3,410 &      .35 &      36,677 &        23 &\clR 11,846 &\clr    19 \\
LJF$_{\lambda=1.0}$    &  1,535,835 &     1,531,663 &\clr  4,172 &      .27 &      29,797 &        15 &\clr 11,084 &        13 \\
RR                     &  2,860,301 &     2,855,350 &      4,951 &      .17 &      24,206 &\clG    10 &     10,305 &         9 \\
SJF$_{\lambda=1.0}$    &  4,334,228 &\clr 4,329,712 &      4,516 &\clr  .10 &      24,578 &        12 &     10,740 &        11 \\
SJF$_{\lambda=0.5}$    &  5,661,022 &\clR 5,657,123 &      3,899 &\clR  .07 &      26,591 &        17 &     11,357 &        15 \\
CR$_{\lambda=0.0}$     &  2,395,472 &     2,381,306 &\clg 14,166 &      .59 &\clg   9,734 &\clg    11 &\clg  1,090 &\clg     7 \\
CR$_{\lambda=0.5}$     &  2,645,302 &     2,635,132 &     10,170 &      .38 &      16,979 &\clG    10 &      5,086 &         8 \\
DJ                     &  1,986,100 &     1,970,895 &\clG 15,205 &      .77 &\clG   2,449 &        26 &\clG     51 &\clG     5 \\
TTL$_{{max}=32}$       &    734,555 &       722,749 &     11,806 &     1.61 &      26,908 &        32 &      3,450 &\clR    20 \\
TTL$_{{max}=32,reset}$ &    740,847 &       728,559 &     12,288 &     1.66 &      23,283 &        32 &      2,968 &\clR    20 \\
TTL$_{{max}=64}$       &    404,840 &       393,213 &     11,627 &     2.87 &      39,416 &        64 &      3,629 &\clr    19 \\
TTL$_{{max}=128}$      &    245,192 &\clg   233,635 &     11,557 &     4.71 &\clr  57,970 &\clr   127 &      3,699 &\clr    19 \\
TTL$_{{max}=128,reset}$&    252,483 &       240,387 &     12,096 &\clg 4.79 &      48,981 &\clr   127 &      3,160 &\clr    19 \\
TTL$_{{max}=256}$      &    165,681 &\clG   154,191 &     11,490 &\clG 6.94 &\clR  86,713 &\clR   255 &      3,761 &\clr    19  
\end{tabular}
\end{table}

By looking on the results for the most restrictive execution time limit of 10~seconds in Table~\ref{tab:10sec}, we observe that even an optimal scheduling algorithm is not able to detect all result updates in the dataset anymore: the
	CV
policy misses 722 query updates.

	LJF
closely outperforms RR regarding update detection. RR again has the smallest maximum delay per query. SJF
	is worse than both LJF and RR in all aspects.

The change-based policy (CR) detects updates more effectively. Without decay ($\lambda=0.0$) the problem occurs, that queries that did not change so far, are executed very rarely. This results in high delays. Since the maximum miss is relatively high and the total miss is low, we infer that only a small number of frequently changing queries is affected. 

The dynamicity-based policy (DJ)
	detects relatively many updates without executing too many irrelevant queries and, thus, is most effective for the scarce time limitation.
Nevertheless, this policy is not starvation-free; it
ignores queries with less updates. Due to the low dynamicity measure they reach at some point, they henceforth receive a very low rank and are not executed anymore. In contrast, queries with more
	frequently changing results
are preferred and get executed repeatedly. The policy actually only selected 6,282 queries\footnote{The number of distinct executed queries is not shown in the table, since it is usually 10,000 for all policies except CV.} from the query set in total, which indicates a cold start problem. As a result, both the maximum delay and the maximum miss grow significantly.

The TTL policies present higher detection rates for short runtime limitations as well. The maximum delay grows with the maximum time-to-live and the configuration TTL$_{{max}=32,reset}$ shows the lowest total delay. It can be seen that more changes are detected with a larger time-to-live, but this comes at the cost of delayed update recognition.
It has to be noted, that the maximum numbers of missed updates are low for all TTL configurations compared to the other policies, even though the delay increases.

\begin{table}[ht]
\centering
\small
\caption{Config $K_\mathsf{maxExecTime}=10sec$}
\label{tab:10sec}
\begin{tabular}{l||r|r|r|r||r|r||r|r}
                       & \begin{tabular}{@{}c@{}} total qe \end{tabular}
                       & \begin{tabular}{@{}c@{}} irrelevant \end{tabular}
                       & \begin{tabular}{@{}c@{}} relevant \end{tabular}
                       & \begin{tabular}{@{}c@{}} eff. (\%) \end{tabular}
                       & \begin{tabular}{@{}c@{}} abs\\ delay \end{tabular}
                       & \begin{tabular}{@{}c@{}} max\\ delay \end{tabular}
                       & \begin{tabular}{@{}c@{}} abs\\ miss \end{tabular} 
                       & \begin{tabular}{@{}c@{}} max\\ miss \end{tabular} \\ \hline
CV                     &    14,484 &           0 &     14,484 &      100 &       2,481 &         2 &        772 &        2 \\ \hline
LJF$_{\lambda=0.5}$    &   690,086 &     687,542 &      2,544 &      .37 &      45,619 &\clg    35 &     12,712 &       31 \\
LJF$_{\lambda=1.0}$    &   780,738 &     778,204 &      2,534 &      .32 &      43,750 &\clG    31 &     12,722 &       28 \\
RR                     &   865,105 &     862,632 &      2,473 &      .29 &      43,097 &\clG    31 &     12,783 &       28 \\
SJF$_{\lambda=1.0}$    &   934,182 &\clr 931,795 &\clr  2,387 &\clr  .26 &      43,681 &        36 &\clr 12,869 &       31 \\
SJF$_{\lambda=0.5}$    & 1,001,825 &\clR 999,526 &\clR  2,299 &\clR  .23 &      43,498 &        38 &\clR 12,957 &       34 \\
CR$_{\lambda=0.0}$     &   109,715 &\clg  99,791 &\clG  9,924 &\clg 9.05 &\clr 152,346 &\clr   678 &\clG  5,332 &\clr  210 \\
CR$_{\lambda=0.5}$     &   676,868 &     671,640 &      5,228 &      .77 &      45,489 &        58 &     10,028 &       46 \\
DJ                     &    17,519 &\clG  11,363 &      6,156 &\clG 35.1 &\clR 499,860 &\clR 2,206 &      9,100 &\clR 1750 \\
TTL$_{{max}=32}$       &   621,510 &     615,662 &      5,848 &      .94 &\clg  37,332 &        39 &      9,408 &\clG   15 \\
TTL$_{{max}=32,reset}$ &   621,250 &     615,380 &      5,870 &      .94 &\clG  34,097 &        38 &      9,386 &\clG   15 \\
TTL$_{{max}=64}$       &   375,209 &     366,929 &      8,280 &     2.21 &      45,342 &        67 &      6,976 &       18 \\
TTL$_{{max}=128}$      &   231,409 &     222,265 &      9,144 &     3.95 &      61,796 &       131 &      6,079 &       18 \\
TTL$_{{max}=128,reset}$&   236,734 &     227,531 &      9,203 &     3.89 &      57,172 &       130 &      6,053 &\clg   16 \\
TTL$_{{max}=256}$      &   162,407 &     152,767 &\clg  9,640 &     5.94 &      95,893 &       258 &\clg  5,574 &       18
\end{tabular}
\end{table}

\section{Conclusions}
\label{sec:conclusions}

	This paper investigates 
multiple performance metrics of scheduling strategies for the re-execution of queries on a dynamic dataset.
	The experiments use
query results gathered from a large corpus of SPARQL queries executed at more than 2,000 time points of the DBpedia~Live dataset, which covers a
period of three months.
The data
	collected
in the experiments has been made public for comparison with other scheduling~approaches.

From the experimental results we conclude that there is no absolute winner.
The execution-time-based policies, \emph{Longest-Job-First} and \emph{Shortest-Job-First}, are not able to compete. Compared to \emph{Round-Robin} they generally perform worse. The main advantage of \emph{Round-Robin}, besides its simplicity, is the constantly short maximum delay, but in any setting it can not convince regarding total delay and change detection.
\emph{Change-Rate} is able to detect a fair amount of changes. An aging factor should be used under scarce execution time restrictions to prevent long delays.
Assuming a limited execution time, the \emph{Dynamics-Jaccard} policy shows best change recognition rates. The effectiveness of this policy as shown in prior work can be confirmed by our results. But, as the execution time limit becomes shorter, this policy tends to disregard queries with low update frequencies. Therefore, it is also not starvation-free. 
As Dividino~\cite{dividino-r-2015-strategies} considered only four iterations, the update frequency of less frequently updated resources could not be measured, but is likely to happen in the dataset update scenario as well.
The \emph{Time-To-Live} policy shows a good performance for update detection and can be well adjusted to a certain maximum delay. It keeps the number of maximum missed changes constant. The alternative configuration to reset the time-to-live value instead of dividing it in half when a change has been detected, proves a better performance and results in higher detection rates and also in reduced delays.

It could be shown, that scheduling strategies based on previously observed changes produce better predictions. The \emph{Time-To-Live} policy can be well adapted to required response times. While the \emph{Change-Rate} and \emph{Dynamics} policies proved to detect most updates, they tend to neglect less frequently changing queries.
Given a less strict execution time limit, \emph{Dynamics-Jaccard} is the best candidate, else \emph{Time-To-Live} can be recommended because it is starvation-free. For future applications it seems reasonable to combine these scheduling approaches into a hybrid scheduler.

\section{Outlook and Future Work}
\label{sec:outlook}

In future work we plan to apply the gathered insights as part of a notification service for query result changes.
To improve the selection of queries in the early stage, we will analyze how change characteristics can be estimated from a~priori knowledge. We observed that the change history is an influential factor for the scheduling strategies. This brings difficulties such as the cold start problem.
To investigate whether the change characteristics of a query can be retrieved from a~priori knowledge~(such as the query itself and its initial execution) we conducted a preliminary analysis: We computed correlations of the query's change probability with different query characteristics, including \emph{query type}, \emph{ordering}, \emph{result limit} and \emph{offset}, \emph{number of result variables}, \emph{number of triple patterns}, \emph{run time} and \emph{result size} at initial execution. Though, no significant correlation could be identified from these features. It will need a deeper examination whether and how the change probability can be predicted from such query characteristics.


\section*{Acknowledgments}\label{sec:acknowledgements}

This work was funded by grants from the German Government, Federal Ministry of Education and Research for the project D-Werft (03WKCJ4D).

{\footnotesize
\bibliographystyle{splncs03}
\enlargethispage{1.0cm}
\bibliography{bibliography}}

\end{document}